\documentstyle[12pt,preprint]{aastex}  

\def\mathnew{\mathsurround=0pt}   
\def\simov#1#2{\lower .5pt\vbox{\baselineskip0pt  
    \lineskip-.5pt\ialign{$\mathnew#1\hfil##\hfil$\crcr#2\crcr\sim\crcr}}}

\def\'#1{\ifx#1i{\accent"13\i}\else{\accent"13#1}\fi}

\begin{document}    
\shorttitle{Non-spherical Structure of the Galactic Dark Matter Halo:
  Stellar Kinematics} \shortauthors{Rojas-Ni\~no et al.}

\title{Detecting Triaxiality in the Galactic Dark Matter Halo through
  Stellar Kinematics}

\author{Armando Rojas-Ni\~no$^{1}$, Octavio Valenzuela$^{1}$, Barbara
  Pichardo$^{1}$, Luis A. Aguilar$^{2}$}     
 
\affil{$^{1}$Instituto de Astronom\'ia, Universidad Nacional
  Aut\'onoma de M\'exico, A.P. 70-264, 04510, M\'exico, D.F.;
  Universitaria, D.F., M\'exico; \\ $^{2}$Observatorio Astron\'omico
  Nacional, Universidad Nacional Aut\'onoma de M\'exico, Apdo. postal
  877, 22800 Ensenada, M\'exico; \\ octavio@astro.unam.mx,barbara@astro.unam.mx}

\begin{abstract} 

Assuming the dark matter halo of the Milky Way as a non-spherical
potential (i.e. triaxial, prolate, oblate), we show how the assembling
process of the Milky Way halo, may have left long lasting stellar halo
kinematic fossils only due to the shape of the dark matter halo. In
contrast with tidal streams, associated with recent satellite
accretion events, these stellar kinematic groups will typically show
inhomogeneous chemical and stellar population properties. However,
they may be dominated by a single accretion event for certain mass
assembling histories.  If the detection of these peculiar kinematic
stellar groups is confirmed, they would be the smoking gun for the
predicted triaxiality of dark halos in cosmological galaxy formation
scenarios.
\end{abstract}


\section{Introduction}                                                     
\label{sec:intro}    
The standard cosmological model ($\Lambda$CDM) has reached a
development stage in which cosmological tests at the scale of galaxies
become possible
\citep[e.g.][]{Bosch1998,Courteau1999,Gnedin2006,Pizagno2007}.
However, a lack of understanding in the complicated physics that
mediates the evolution of baryonic matter has meant that these tests
are hard to implement, as their application depends on this
physics. In some cases, controversial results are found out of these
studies \citep[e.g.][]{Klypin1999,Moore1994}, stimulating suggestions
of modifications for the model, but probably also the need of more
accurate comparisons between theory and observations
\citep[e.g.][]{Valenzuela2007,Simon2007}.

Under the $\Lambda$CDM picture of structure formation, galactic sized
halos are assembled by the accretion of smaller structures or subhalos
\citep{Kauffmann1993,Ghigna1998,Klypin1999}, or by major mergers,
leaving imprints from these processes in the dark matter kinematics
and leading to a triaxial halo shape.  Besides the assembling process,
dark matter halos triaxiality may be produced by dynamical secular
evolution like the radial orbit instability
\citep{Aguilar1990,Barnes2005}. All together triaxiality is an
ubiquitous property of dark matter structures in a hierarchical
universe \citep{Allgood2006,Vera2011}. Studies aiming to detect
triaxiality, have to face the complication of halo shape evolution or
even that triaxialty may be erased or turned into oblateness by the
baryonic galaxy formation, as a consequence, a robust detection or
rejection of a non-spherical halo is still a valuable result
constraining the efficiency of dynamical mechanisms of halo
triaxiality evolution.  Different observational strategies have been
proposed in order to quantify triaxiality in galactic halos.  One of
them is based on the Sagittarius stellar tidal stream morphology and
kinematics \citep{Law2009}, other on hypervelocity stars
\citep{Gnedin2005}, one more on galaxy satellite systems distribution
\citep{Zentner2005}. However, a clear observational detection of dark
matter halo non-sphericity is still a challenge, mostly because of the
need of higher precision in observations and because of degeneracies
triggered by the baryonic galaxy structure. Based on the previous
discussion our paper is focused on detecting a non-spherical halo,
therefore we will use triaxiality in a very broad sense without
distinguishing between true triaxiality, prolateness or oblateness.
However at the end of the paper we will briefly discuss possible
differences.

The aim of this paper is to introduce a possible new strategy in order
to detect triaxiality in the Milky Way halo.  Our strategy is a
generalization of the discussion presented by \citet{Penarrubia2009},
for dwarf spheroidal satellite galaxies. \citet{Penarrubia2009} shows
that globular clusters stellar remnants are long lived in the
potential of triaxial dwarf galaxies, explaining the origin of cold
stellar structures observed in UMinor and other Milky Way
satellites. The kinematic groups correspond to halo version of
galactic disks stellar moving groups \citep{Antoja09}, however the
mechanism is not identical.  In this work we propose that if the Milky
Way sits in a triaxial dark matter halo, the stellar halo assembling
history is able to populate the quasi-resonant orbits triggering
kinematic stellar groups.  In order to prove our thesis, we perform
numerical simulations of test particles moving inside the potential
generated by dark matter halos with different shapes, the population
of the orbital structure is randomly generated.  We analyze the
resulting kinematic and orbital structure in order to decide if the
stellar kinematic structure may be evidence for the halo triaxiality.

This letter is organized as follows. In Section \ref{model} the 3-D
galactic halo potential used to compute orbits is briefly
described. In Section \ref{simulations} we introduce a strategy aimed
to efficiently explore the stellar phase space accessible to an
hypothetical observer, we also present the results of our numerical
simulations. Finally, in Section \ref{conclusions} we present a
discussion of our results and our conclusions.

\section{The Model for the Galactic Halo}\label{model} 

For simplicity our model assumes that the Galaxy is surrounded by a
static dark matter halo with the density profile proposed by
\citet{Navarro1996}.  It is important to mention that in our study we
do not pretend to use a Milky Way state of the art model, instead we
present a proof of concept that a triaxial halo develops and preserves
abundance structure in the stellar phase space because of the resonant
orbital structure.  For example, we do not include yet any disk effect
even that Solar Neighborhood stellar orbits will be clearly
influenced. Even more, the Galactic dark matter halo shape has been
evolving during time, however we still use a static model. Our study
does not loose generality because, as we show, the dominant halo
resonant orbital structure is present only inside a non-spherical
halo.  In any case, our results provide a lower limit to the possible
galactic triaxiality, which may be comparable to the kinematics
sampled by an observer further away from the disk or after discarding
if possible, any disk effect.  We will return to this point in the
Discussion section.

Under our assumptions, the gravitational potential generated by the
halo takes the form

\begin{equation}
  \Phi {\rm (x,y,z)}=2\pi G abc\rho_0 r_s^2 \int_0^\infty \frac
       {s(\tau)}{r_s+s(\tau)} \frac{d\tau } {\sqrt
         {(a^2+\tau)(b^2+\tau)(c^2+\tau)}},
\label{phieq}
\end{equation}
from \citet{Penarrubia2009}. Here, the dimensionless quantities $a$,
$b$ and $c$, are the three main axes, $\rho_0$, is the
characteristic density of the halo and $r_S$, is the radial scale. We
use elliptical coordinates, where,

\begin{equation}
  s(\tau)=\frac{x^2} {a^2+\tau}+\frac{y^2} {b^2+\tau}+\frac{z^2} {c^2+\tau}.
\label{elipeq}
\end{equation}
Halo triaxiality is given then, by $a$, $b$ and $c$. 

In the next section, we explore the effect of halo triaxiality in the
resulting orbital structure. The model density structure is determined
by five free parameters, $r_S$, $\rho_0$ and the axis ratios a, b and
c. Ideally, all can be adjusted using Milky Way observations, however
this is a not a finished task in Galactic astronomy and we only seek to 
distinguish between orbits in a spherical and triaxial halo. Instead we adopt
$r_S=8.5$ kpc, in order to guarantee a flat rotation curve at the
solar radius in the absence of a disk, and $\rho_0$= 2.46 M$\odot$
pc$^{-3}$, obtained assuming a maximum rotation velocity of 220 km/s.
The possible axes ratios are even harder to constrain, thus, we
decided to explore four cases: One (a=1.47, b=1.22, c=0.98),
motivated by the halo model that best fits Sagittarius stream
\citep{Law2009}, a prolate  model, an oblate one and a spherical halo model.  

\section{Numerical Test Particle Simulations}\label{simulations} 
We tracked the orbital motion of a number of stars under the influence
of the gravitational potential generated by a steady triaxial dark
matter halo as described in the previous section. The equations of
motion are numerically solved with the Bulirsh-Stoer method
\citep{Press1992}. We monitored the individual particle energy
constant of motion and obtain a maximum relative error in our
computations of about $10^{-9}$.  We will analyze two different set
ups that will enable us to demonstrate the power of orbital structure
in order to identify a non-spherical halo.  After 12 Gigayears , which
is comparable to the stellar halo formation timescale
\citep{Kalirai2012}, we analyzed the kinematic projections vx-vz and
we use the spectral orbital method developed by \citet{Carpintero1998}
in order to study the orbital type distribution and investigate if the
structure is composed of resonant orbits. This method is based on the
concept of spectral dynamics introduced by Binney \& Spergel (1984)
that uses the Fourier transform of the time series of each
coordinate. The method distinguishes correctly between regular and
irregular orbits, and identifies various families of regular orbits
(boxes, loops, tubes, boxlets, etc.), it also recognizes the
second-rank resonances that bifurcate from them.

\subsection{Satellite Infall Initial Conditions}\label{satellite} 
\begin{figure}
\includegraphics[width=0.5\textwidth]{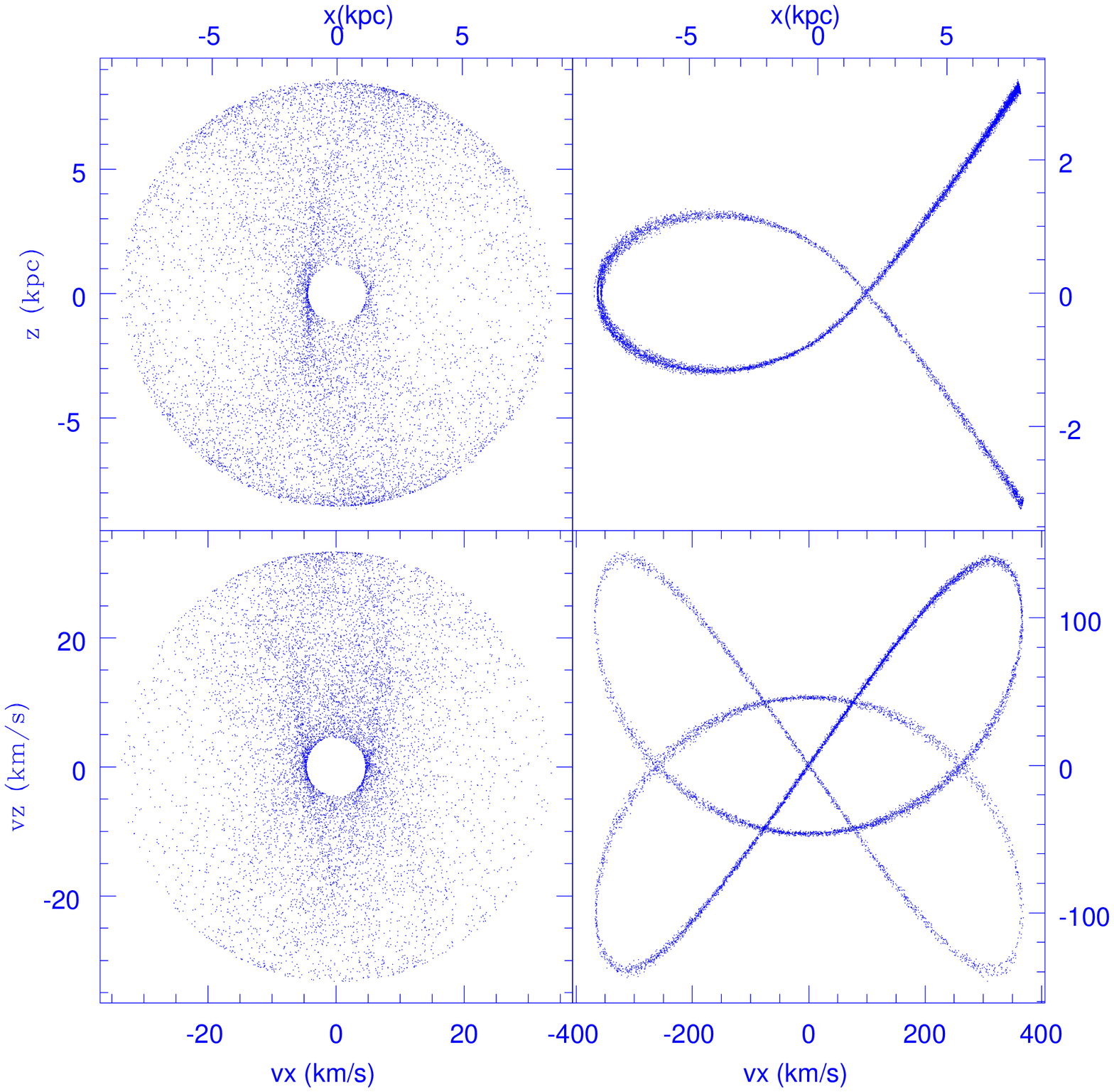}
\caption{Position and velocities of the satellite stars after 16
  Gigayears for the spherical (left) and the triaxial halo (right). It
  is remarkable the small mixing shown after the long integration time
  for particles close to the 2:3 resonant orbit in the triaxial case.}
\label{fig_posvel}
\end{figure}

Our first set of simulations study the infall of a satellite or
globular cluster into the dark matter halo potential, near to a
resonant orbit (2:3). An important characteristic of a triaxial
potential is that resonant orbits are dominant compared to the
situation in a spherical potential where there is no angular
preference for particles to settle down, and the abundance of all type
of orbits is not dominated by resonances.  This circumstance is an
important ingredient to determine the halo triaxiality.  The aim of
this set of controlled experiments is to show that in the vicinity of
resonances, stellar orbits librate and stay for practically infinite
time. In our simulation, the satellite is made of $10^{4}$ test
particles with an isotropic velocity dispersion of 10 km/s, the number
of particles has no effect on the orbital calculation accuracy,
however it may impact the significance of kinematic structures, we
will comeback to this issue later. Satellite self-gravity is
neglected, however including this effect will produce only a delay in
the satellite disruption time and little or no effect in the abundance
of kinematic stellar structure.  We let the system evolve for a period
comparable to recent constraints on the stellar halo formation time
(12 Gigayears), under the potential generated by the dark halo for the
two cases mentioned in Section \ref{model}. Figure \ref{fig_posvel}
shows the resultant position and velocities for the satellite stars.

\begin{figure}
\includegraphics[width=0.5\textwidth]{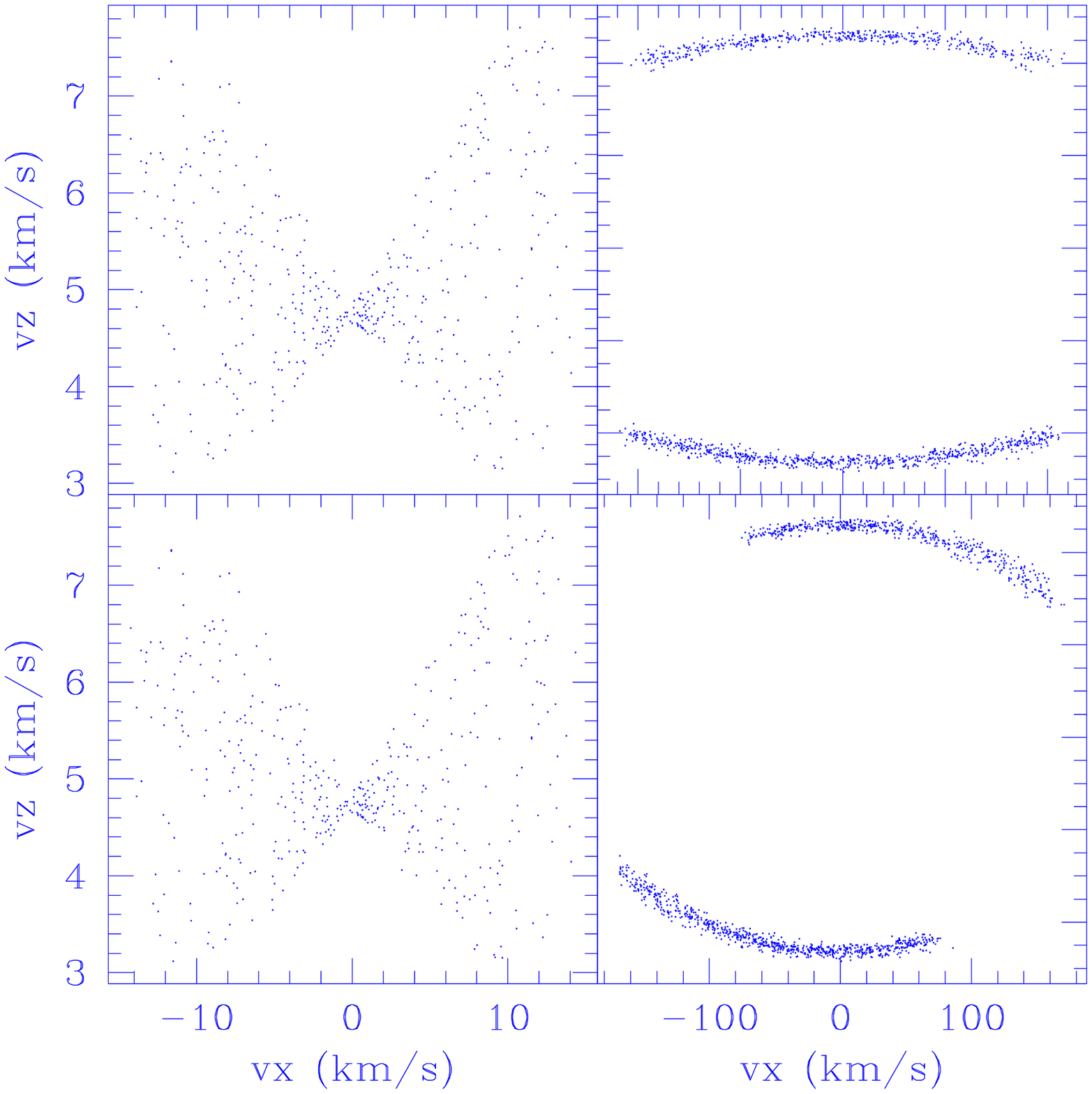}
\caption{Velocity space in the solar neighborhood (upper panels) for
  both cases spherical (left panel), triaxial (right panel), setting
  the Sun close to the fish head orbit. Bottom panels show the case
  for an observer further away by 20 kpc for both cases spherical
  (left panel), triaxial (right panel).}
\label{fig_resvel}
\end{figure}

In Figure \ref{fig_posvel} (upper panels), the particle positions
indicate that disrupted stars with the right energy and angular
momentum librate around a resonant orbit during all the simulation
time for the triaxial halo. In contrast, for the spherical case it is
clear that particles are considerably more mixed by the end of the
simulation. Figure \ref{fig_posvel} (lower panels) shows how the stars
in the vicinity of resonances are prone to form kinematic groups for
the triaxial case in contrast to what it happens in the spherical
halo.  This is similar to what has been recently claimed by
\citet{Lux2012} for a galactic globular cluster. These groups would
look different from different positions with respect to this fish-like
orbit although we have focused on the difference between the spherical
and the triaxial case. Figure \ref{fig_resvel} (upper panels) shows
how velocity space would look for example at the solar neighborhood at
the end of the simulation, if the Sun were close to the ``fish head''
orbit (left panels for a spherical halo and right panels for a
triaxial halo). The observer will find two symmetric arm-like
kinematic structures close to 100 km/s. We assumed the solar position
at (x=-8.5 kpc, y=0, z=0), and take as stars in the ``solar
neighborhood'', those found inside a sphere of radius 1 kpc, centered
at the solar position, however we could have placed the observer at a
different position. In order to illustrate that, Fig. \ref{fig_resvel}
(bottom panels) shows also the velocity space, but for stars that are
inside a sphere centered at (x=-8.5 kpc, y=0, z=20 kpc).

For both cases, although at different positions, the velocity space
shows two arms that correspond to stars that are moving along the
resonant orbit in opposite directions. Kinematic groups formed in this
manner would provide an evidence of a non-spherical halo.

\subsection{Random Velocity Initial Conditions}\label{satellite} 
Once we provided a good evidence of kinematic stellar structure
triggered by resonances in a triaxial dark matter halo in a single
satellite accretion event, we have to asses the probability that a
long stellar halo assembly history may populate orbits in the vicinity
of resonances triggering long lasting stellar moving groups. We face
three difficulties, the number of particles is inversely proportional
to noise which could artificially mimic the kinematic groups,
therefore if we want to have a reliable moving group detection we need
to have a large number of particles. Secondly, an observer will have
good kinematic accuracy only inside a limited neighborhood, finally
during the galaxy assembly history, satellites and globular clusters
have been accreted from many different directions. For that reason
instead of populating the whole halo with particles, we use the
following initial conditions set-up.  We define a sphere with radius
R, around a putative observer, we randomly populate the sphere with
particles, also with randomly selected velocities between zero and the
local escape velocity, mimicking the contribution of many accretion
events with different orientations and energies, etc, but assuring
that the corresponding orbit is bounded to the galaxy.  This is just
an strategy to explore the available phase space, is not pretending to
be a full self-consistent generation of position and velocities. Stars
librating around resonances will comeback eventually to the observer
position therefore we populate orbits that the artificial observer
will detect, increasing our particle statistics at a low computing
price.  The process produce persisting kinematic groups at the
velocities around resonances if the process is efficient to explore
the available phase space, because irregular and open orbits will
almost never comeback.  In contrast a non-resonant region will present
nearly evenly distributed types of orbits as a result of phase mixing,
and also non-kinematic groups after some mixing time scales.

\begin{figure}
\includegraphics[width=0.5\textwidth]{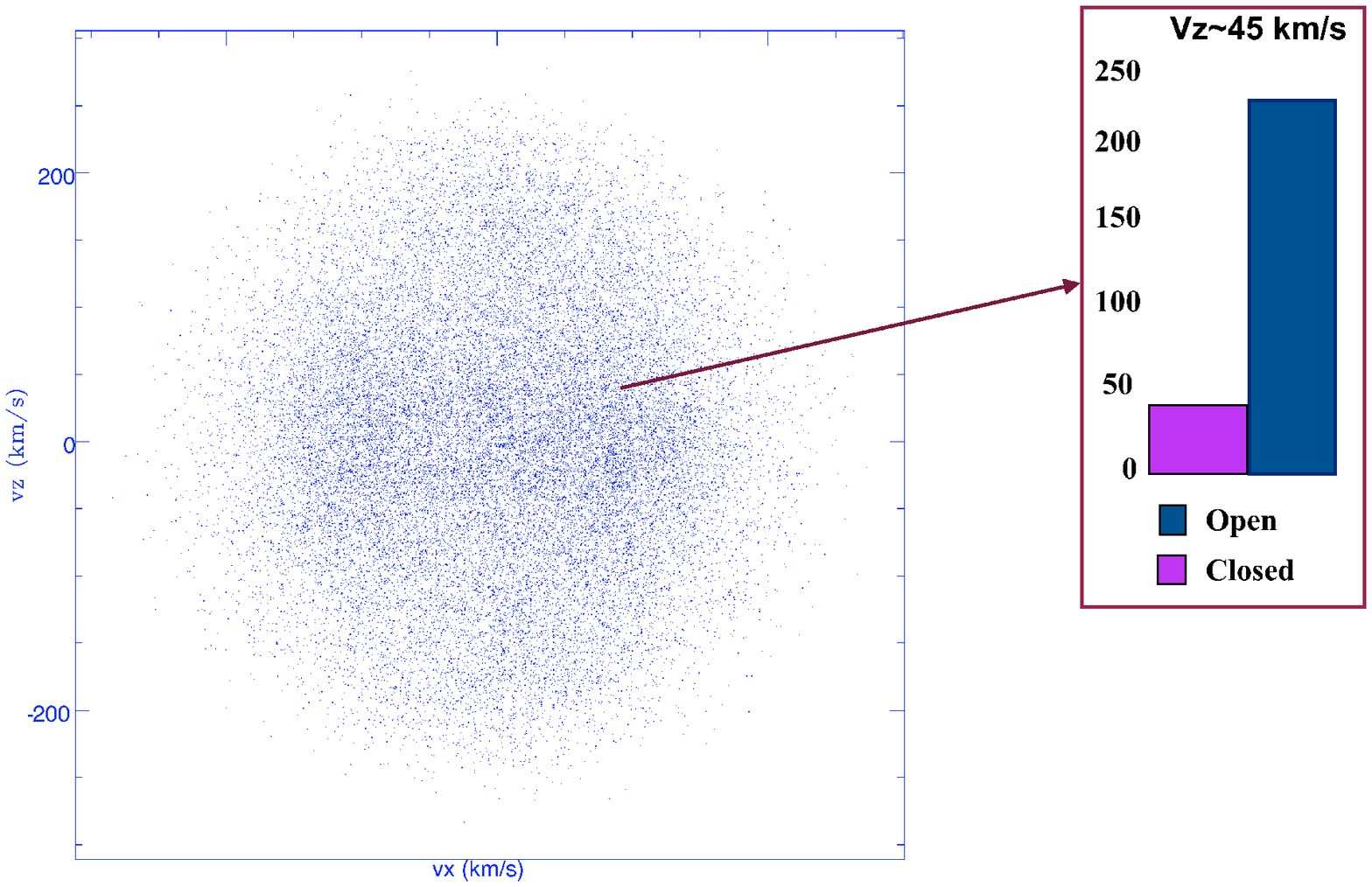}
\includegraphics[width=0.5\textwidth]{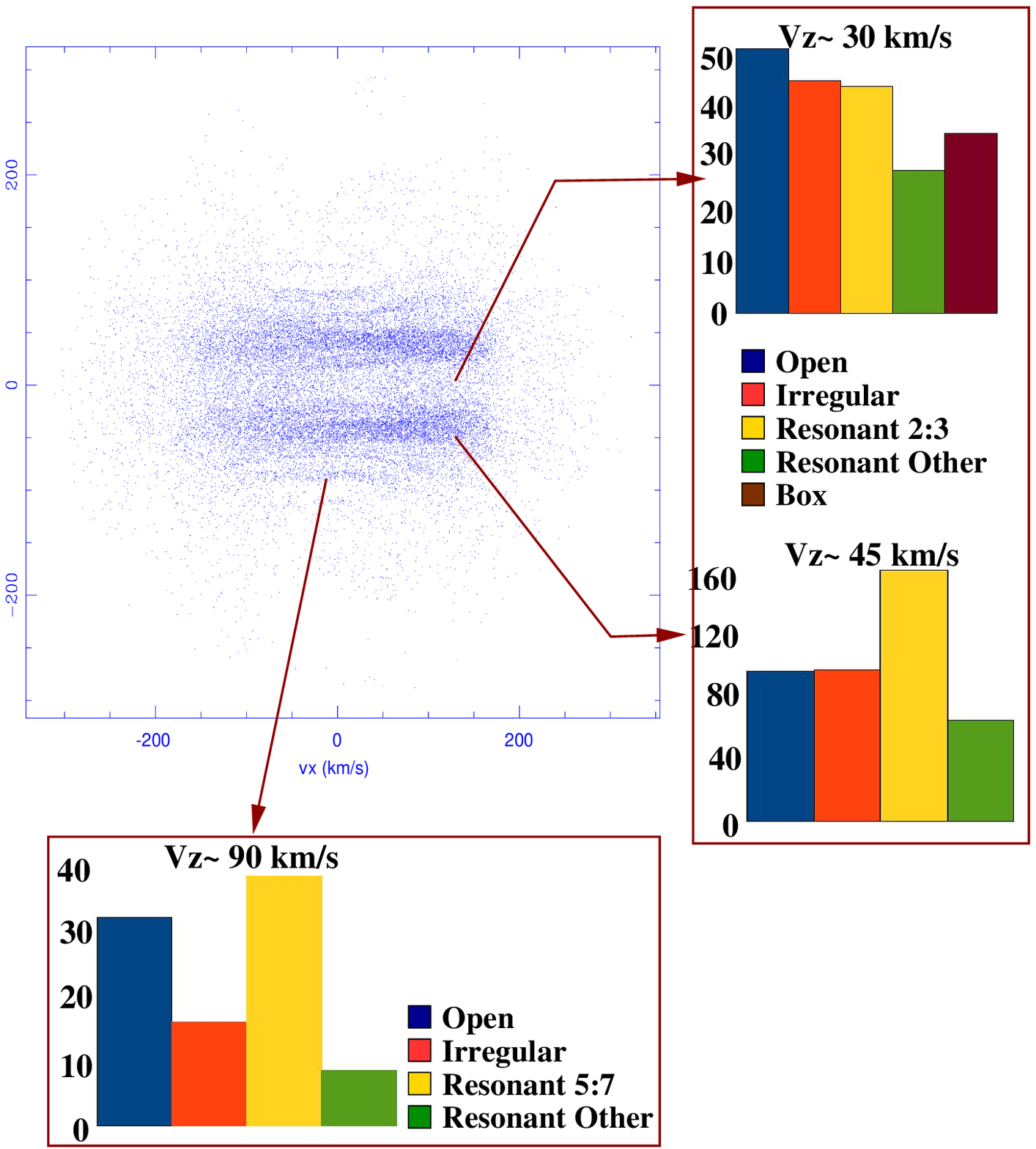}

\caption{Kinematic structure in spherical (top panel) and triaxial
  halo (bottom panel) models. We show the vx-vz projection of the
  velocity space measured by a hypothetical observer inside the
  spherical model, the projection is basically featureless, the inset
  shows a histogram of the orbital types, the contribution of resonant
  orbits is negligible. Right hand panel shows the vx-vz diagram
  corresponding to an observer inside the triaxial halo model. We
  recover the structure related to the fish-like orbits as in the
  single satellite experiment, and other structures, regardless that
  in this case we randomly populated the orbital distribution.  We
  present three insets in the triaxial halo case: the upper right one
  that shows almost a nearly even distribution of all orbital species,
  the inset in the middle shows a distribution clearly dominated by
  resonant orbits (2:3), the lower left inset is centered in another
  kinematic structure, again the 5:7 resonant is dominant. We conclude
  that the orbital type distribution supports the resonant origin of
  the kinematic structure and its capability as a triaxiality
  diagnostic. Both systems were evolved for $2 \times 10^9$ years. }
\label{fig_alea}
\end{figure}

We set $2\times 10^6$ stars randomly distributed inside the ``observer
neighborhood'', defined as mentioned above, we assigned random
velocities between zero and the escape velocity.  We removed from the
initial conditions the stars that were moving close to the X-Y
plane. These stars remain in this plane all along the simulation and
will always have vz close to zero. If we have populated the whole halo
this orbits concentration would be compensated with neighbor orbits
thrown out from different galaxy positions.  Although this orbits
would appear in the velocity plane as an horizontal band, they do not
form a kinematic group, we verified this using an spectral orbit
clasification \citep{Carpintero1998}, as discussed below. We let the
system evolve during 12 Gigayears and at the end of this integration
time, we studied the stellar kinematic distribution.  This integration
time is long enough to be comparable to the Milky Way halo evolution
\citep{Kalirai2012}.  For our purpose, we consider the stars that are
found after the simulation, in the ``artificial observer
neighborhood''. Figure \ref{fig_alea} shows the resulting kinematic
distribution of stars after performing the simulation for two
different cases. Upper panel of Fig. \ref{fig_alea} corresponds to a
spheric halo (a=1.0, b=1.0, c=1.0) and bottom panel corresponds to a
triaxial halo (a=1.47, b=1.22, c=0.98).

From Figure \ref{fig_alea}, the velocity space for the triaxial halos
shows two arms in the same position as in the case of the falling
satellite. The fact that the kinematic structures do not appear in the
spheric case strongly suggests that these kinematic arms correspond to
stars librating in the vicinity of a resonance. In order to proof our
interpretation, we use the spectral orbital method mentioned in
Section \ref{simulations} and show that the more outstanding kinematic
structures have an important contribution of resonant orbits, in
contrast a featureless region has an almost even distribution of
orbital types, and the spherical case shows mostly open
orbits. Although we do not pretend to assess yet the detectability of
the kinematic structure triggered by a non-spherical halo we would
like to show that the importance of resonant orbits in our velocity
structure is not an artifact of small number statistics or poor
sampling. We would like to stress that the orbital type histogram
presented in figure 3 shows that particles in the 2:3 resonance are
38.5\% of the total, significantly larger than the typical fluctuation
amplitude.  Similar calculation were performed using an oblate and
prolate models, results although non identical they are qualitatively
the same as the ones presented in this paper. The corresponding
details will be presented in a future study.

\begin{figure}
\includegraphics[width=0.5\textwidth]{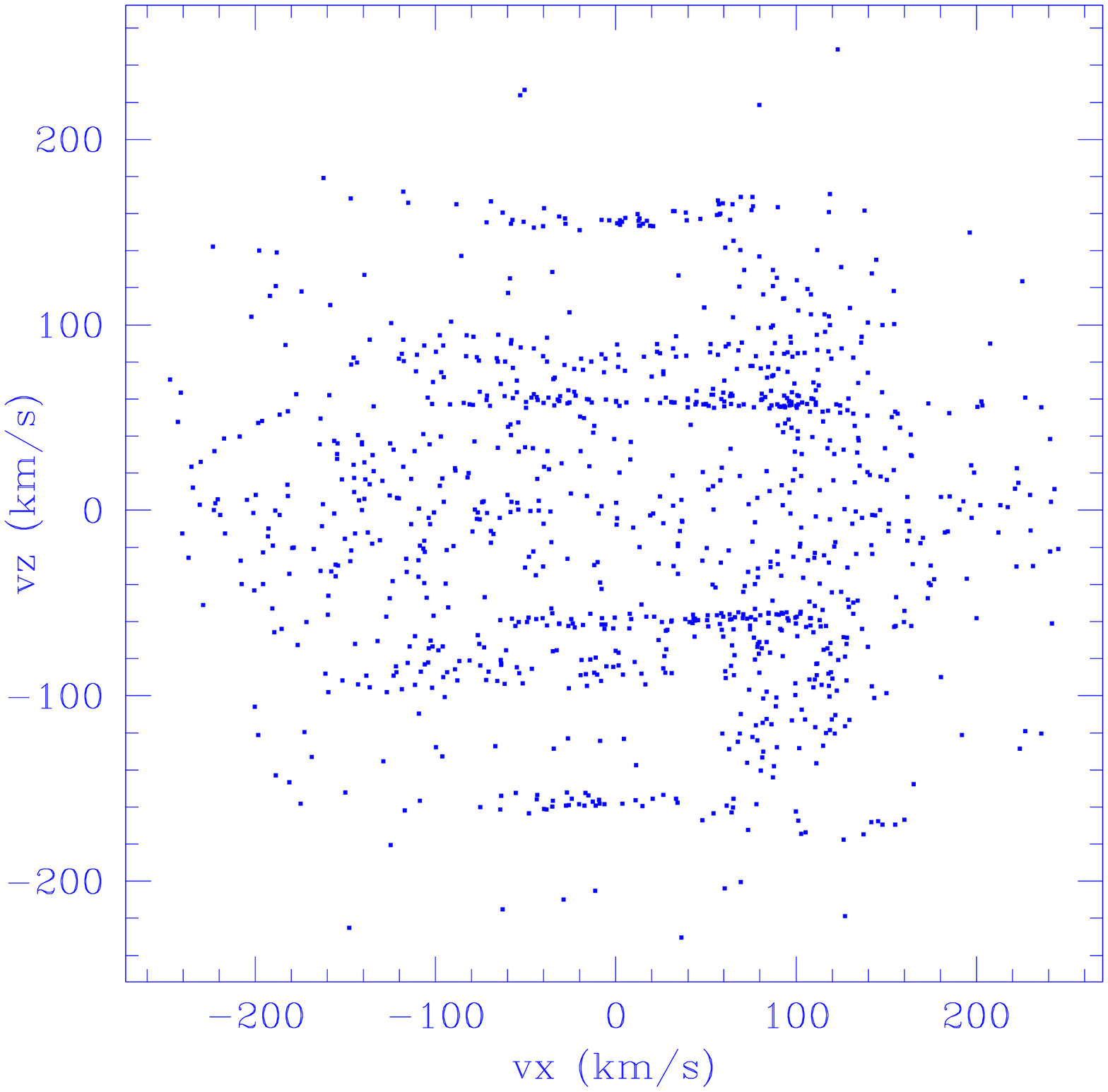}
\caption{Same as Figure \ref{fig_alea} c) and d), but for a different
  position (8.5,0,1.5). Regardless of the observer height with respect
  to the disk plane, it is possible to detect the signature of the non
  spherical halo.}
\label{fig_Zalea}
\end{figure}

Fig. \ref{fig_Zalea} shows also the velocity space for the triaxial
halos, but for stars that are inside a sphere centered at a large
distance from the galactic disk. Notice that the arm-like features are
still detected, just as in the satellites case. Therefore, details of
the observer´s position and galactic model, through important are not
key in order to detect the kinematic groups.

\section{Discussion and Conclusions}\label{conclusions} 
Using test particle simulations, we studied the motion of stars under
the influence of the gravitational potential generated by a steady
triaxial dark matter halo and a spherical one. Our simulations show
that a non-spherical shape of the Milky Way dark matter (triaxial,
prolate, oblate) halo determines an important amount of the stellar
halo kinematic structure, and that this is a long lasting
feature. There is some dependence on the kinematic structure
distinguishing the triaxial case from the prolate and oblate
situation, however the most outstanding difference is the comparison
with the spherical case, where no kinematic structure is triggered.

In a satellite accretion event into the triaxial halo potential, close
to a resonant orbit, star particles librate around resonances for
times longer than the Universe age (16 Gyrs).  Even if the stars have
initial random velocities, mimicking random satellite accretion events
during the Galactic halo assembling, many of the stars still stay
around resonant orbits and are prone to form kinematic groups if the
dark matter halo is triaxial.  The reason is that an important
fraction of stars will be librating around resonances and they will be
prone to form kinematic groups (See Figs. \ref{fig_alea} and
\ref{fig_Zalea}). These groups might be observationally detectable,
however this endeavor will face important challenges.  First, the disk
contribution to the gravitational potential may trigger orbital
structure as it has been discussed by \citet{Lux2012}, however either
the disk effect may be characterized and avoided, or the search may
focuse on stars further away from the disk effect in order to minimize
the disk graviational effect.  The era of GAIA will provide us with
strong constraints to the Galaxy stellar structure, as well as 6D
spatial information and chemical composition of a huge number of stars
and at distances further away from the solar neighborhood. Careful
analysis of these data compared with the results of our simulations
may shed some light on the shape of the Milky Way halo. A single event
such as the tidal disruption of a globular cluster or a satellite, may
produce long lasting kinematic groups even in configuration space
\citep{Lux2012, Penarrubia2009} with some degeneracy in their
interpretation due to the dark matter halo and the dynamical age of
the accretion event.  However, we will be able to distinguish these
groups from those associated with the dark matter halo shape with the
help of the stellar population data, which must be very homogeneous.
The detection of these type of kinematic groups, would be a clear
indication of the triaxiality (non-sphericity) of the Milky Way dark
matter halo, and will be of great importance for the galaxy formation
theories and the $\Lambda$CDM scenario.

\acknowledgments We thank Antonio Peimbert for some enlightening
discussions. We thank DGAPA-PAPIIT through grants IN110711, IN117111-3 
and CONACyT grant CB-128556.


\begin{thebibliography} 

\bibitem[Aguilar \& Merritt(1990)]{Aguilar1990} Aguilar, L.~A., \&
  Merritt, D.\ 1990, \apj, 354, 33

\bibitem[Allgood et al.(2006)]{Allgood2006} Allgood, B., Flores, 
R.~A., Primack, J.~R., et al.\ 2006, \mnras, 367, 1781 

\bibitem[Antoja et al.(2009)]{Antoja09} Antoja, T., Valenzuela, 
O., Pichardo, B., et al.\ 2009, \apjl, 700, L78 

\bibitem[Barnes et al.(2005)]{Barnes2005} Barnes, E.~I., Williams, 
L.~L.~R., Babul, A., \& Dalcanton, J.~J.\ 2005, \apj, 634, 775 

\bibitem[Binney \& Spergel(1984)]{Binney1984} Binney, J., \& Spergel,
  D.\ 1984, \mnras, 206, 159

\bibitem[Carpintero \& Aguilar(1998)]{Carpintero1998} Carpintero,
  D.~D., \& Aguilar, L.~A.\ 1998, \mnras, 298, 1

\bibitem[Courteau \& Rix(1999)]{Courteau1999} Courteau, S., \& Rix,
  H.-W. 1999, \apj, 513, 561

\bibitem[Ghigna et al.(1998)]{Ghigna1998} Ghigna, S., Moore, B.,
  Governato, F., Lake, G., Quinn, T., \& Stadel, J. 1998, \mnras, 300,
  146

\bibitem[Gnedin et al.(2005)]{Gnedin2005} Gnedin, O.~Y., Gould, 
A., Miralda-Escud{\'e}, J., \& Zentner, A.~R.\ 2005, \apj, 634, 344 

\bibitem[Gnedin et al.(2006)]{Gnedin2006} Gnedin, O., Weinberg, D. H.,
  Pizagno, J., Prada, F., Rix, H. W. 2006, a-ph0607394

\bibitem[Jing \& Suto(2002)]{Jing2002} Jing, Y.~P., \& Suto, Y.\ 2002,
  \apj, 574, 538


\bibitem[Kalirai(2012)]{Kalirai2012} Kalirai, J.~S.\ 2012, \nat, 
486, 90 

\bibitem[Kauffmann et al.(1993)]{Kauffmann1993} Kauffmann, G., White,
  S. D. M., \& Guideroni, B. 1993, \mnras, 264, 201 Klypin, A.A.,
  Kravtsov, A.V., Valenzuela, O., \& Prada, F. 1999, \apj, 522, 82

\bibitem[Klypin et al.(1999)]{Klypin1999} Klypin, A., Gottl{\"o}ber,
  S., Kravtsov, A. V., \& Khokhlov, A. M. 1999, \apj, 516, 530

\bibitem[Law et al.(2009)]{Law2009} Law, D.~R., Majewski, S.~R., \&
  Johnston, K.~V.\ 2009, \apjl, 703, L67

\bibitem[Lux et al.(2012)]{Lux2012} Lux, H., Read, J.~I., Lake, G., \&
  Johnston, K.~V.\ 2012, \mnras, L464

\bibitem[Moore(1994)]{Moore1994} Moore, B. 1994, Nature, 370, 629

\bibitem[Navarro et al.(1996)]{Navarro1996} Navarro, J.~F., Frenk,
  C.~S., \& White, S.~D.~M.\ 1996, \apj, 462, 563

\bibitem[Pe{\~n}arrubia et al.(2009)]{Penarrubia2009} Pe{\~n}arrubia, 
J., Walker, M.~G., \& Gilmore, G.\ 2009, \mnras, 399, 1275

\bibitem[Pizagno et al.(2007)]{Pizagno2007} Pizagno, J., et al. 2007,
  \aj, 134, 945

\bibitem[Press et al.(1992)]{Press1992} Press, W. H., Teukolsky,
  S. A., Vetterling, W. T., \& Flannery, B. P. 1992, Numerical Recipes
  in Fortran 77: The Art of Scientific Computing (2d ed.; Cambridge:
  Cambridge Univ. Press)

\bibitem[Simon \& Geha (2007)]{Simon2007} Simon, J. D., \& Geha,
  M. 2007, ArXiv e-prints, 706, arXiv: 0706.0516

\bibitem[Valenzuela et al.(2007)]{Valenzuela2007} Valenzuela, O.,
  Rhee, G., Klypin, A., Governato, F., Stinson, G., Quinn, T., \&
  Wadsley, J. 2007, \apj, 657, 773

\bibitem[Van den Bosch(1998)]{Bosch1998} Van den Bosch, F. C. 1998,
  \apj, 507, 601

\bibitem[Vera-Ciro et al.(2011)]{Vera2011} Vera-Ciro, C.~A., 
Sales, L.~V., Helmi, A., et al. 2011, \mnras, 416, 1377 

\bibitem[Zentner et al.(2005)]{Zentner2005} Zentner, A. R., Kravtsov,
  A. V., Gnedin, O. Y., \& Klypin, A. A. 2005, \apj, 629, 219


\end{thebibliography}
\end{document}